\documentclass[%
 reprint,
 amsmath,amssymb,
 aps,
]{revtex4-2}

\usepackage{graphicx}%
\usepackage{dcolumn}%
\usepackage{bm}%
\usepackage{hyperref}%
\usepackage{caption}
\usepackage{subcaption}
\usepackage[normalem]{ulem}

\usepackage{booktabs}
\usepackage{cellspace}
\newcommand{\shfx}{\overline{w'\theta'}}

\begin{document}

\title{$\Pi$-ML: A dimensional analysis-based machine learning parameterization \\of optical turbulence in the atmospheric surface layer}

\author{Maximilian Pierzyna}
\email{Corresponding author: m.pierzyna@tudelft.nl}
\affiliation{Faculty of Civil Engineering and Geosciences, Delft University of Technology, Delft, Netherlands }

\author{Rudolf Saathof}
\affiliation{Faculty of Aerospace Engineering, Delft University of Technology, Delft, Netherlands}

\author{Sukanta Basu}
\affiliation{Atmospheric Sciences Research Center, University at Albany, Albany, USA}

\date{\today}%

\begin{abstract}
    Turbulent fluctuations of the atmospheric refraction index, so-called optical turbulence, can significantly distort propagating laser beams. 
    Therefore, modeling the strength of these fluctuations ($C_n^2$) is highly relevant for the successful development and deployment of future free-space optical communication links. 
    In this letter, we propose a physics-informed machine learning (ML) methodology, $\Pi$-ML, based on dimensional analysis and gradient boosting to estimate $C_n^2$.
    Through a systematic feature importance analysis, we identify the normalized variance of potential temperature as the dominating feature for predicting $C_n^2$.
    For statistical robustness, we train an ensemble of models which yields high performance on the out-of-sample data of $R^2=0.958\pm0.001$.
\end{abstract}

\keywords{Optical turbulence, Physics-informed machine learning}%
\maketitle

Free-space optical communication (FSOC) between satellites and ground or between multiple ground terminals is among emerging applications in which an optical beam propagates through the atmosphere.
FSOC can have a major societal impact, increasing data throughput, data security, and global internet coverage while potentially reducing the cost per bit per second \citep{Hemmati2009}.
However, some challenges need to be addressed; besides precipitation, clouds, fog, and aerosol scattering, turbulent fluctuations of the atmospheric refractive index form a major source of disturbance \citep{kaushal2017}.
The strength of these fluctuations -- called optical turbulence -- is quantified by the refractive index structure parameter $C_n^2$.
Good knowledge about the behavior of $C_n^2$ in diverse locations and meteorological conditions is required to design and deploy reliable future FSOC links.
However, measuring $C_n^2$ is difficult and typically needs elaborate post-processing of high-frequency observations \citep{wyngaard1971}.
As a result, a wide range of empirical $C_n^2$ models and parameterizations have emerged, which aim to relate $C_n^2$ to more easily obtainable variables \citep{smith1993}.
Conventional physics-based $C_n^2$ parametrizations typically make use of Monin-Obukhov similarity theory (MOST) \citep{monin1954} and associated empirically determined similarity relationships.
One of the earliest parameterizations was proposed by \citep{wyngaard1971} and utilizes turbulent fluxes to estimate $C_n^2$.
Several other competing formulations exist (refer to \citep{savage2009} for a comprehensive review).

Recently, multiple studies \citep{wang2016,jellen2021,bolbasova2021,su2021} showed that machine learning (ML) models can be used to parametrize $C_n^2$ based on routinely-available meteorological inputs.
These ML approaches parametrize the underlying physical processes from data through sophisticated regression, but they do not explicitly incorporate physical knowledge.
In this letter, we propose an alternative physics-inspired ML framework.
We present $\Pi$-ML, a dimensional analysis-based ML framework, which strives to improve conventional MOST-based surface layer parameterizations with the power of ML.
We utilize dimensional analysis (DA) constrained with domain knowledge to expand the set of traditional MOST variables and
an ensemble of gradient-boosting ML regression models to learn similarity relationships from observations.
In DA, the relevant dimensional variables of a physical process are combined into non-dimensional groups which describe that process equally well  \citep{stull1988}.
DA is compelling to use in practice because the non-dimensional variables enable us to combine observational data from different field campaigns around the world.
More importantly, when using ML, DA can change the extrapolation problem in dimensional variables to an interpolation problem in non-dimensional variables \citep{kashinath2021}.
The $\Pi$-ML code available at \url{https://github.com/mpierzyna/piml}.

\begin{figure*}[t]
    \centering
    \includegraphics[width=\textwidth]{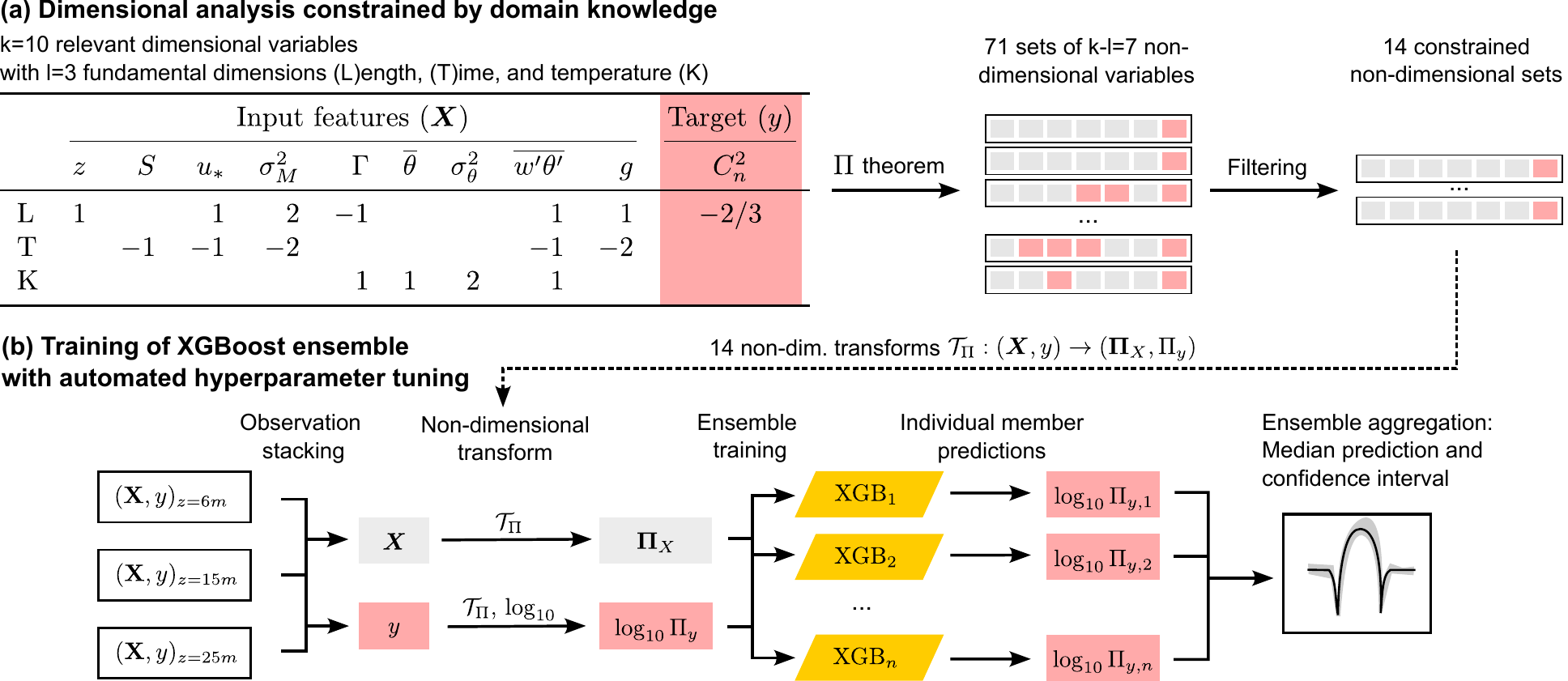}
    \caption{Our $\Pi$-ML methodology consists of two components. (a) The dimensional analysis based on the Buckingham $\Pi$ theorem combines observed dimensional variables into $\Pi$ sets of normalized non-dimensional variables. (b) These sets are used to transform the observed data into a stacked non-dimensional dataset to train an ensemble of XGBoost regression models.}%
    \label{fig:methodology}
\end{figure*}
To investigate the strengths and weaknesses of the proposed methodology, we use measurements collected during a seeing study at the Mauna Loa Observatory (MLO) on the island of Hawai'i.
The MLO study was conducted by the National Center for Atmospheric Research (NCAR) from 9 June 2006 until 8 August 2006 ($\sim$8 weeks).
The publicly available dataset contains measurements of 
mean meteorological quantities, %
turbulent fluxes, %
and turbulent variances %
obtained from three sonic anemometers deployed at ca.\ 6m, 15m, and 25m altitude.
$C_n^2$ values were estimated by NCAR via inertial-range scaling of temperature spectra \citep{oncley2013}.
We compute two gradients from the mean horizontal wind components $\overline{u}$ and $\overline{v}$ and the mean potential temperature $\overline{\theta}$:
mean wind shear $S=\sqrt{{(\partial\overline{u}/\partial z)}^2 + {(\partial\overline{v}/\partial z)}^2}$ 
and mean potential temperature gradient $\Gamma= \partial\overline{\theta}/\partial z$.
Atmospheric turbulence is modulated by thermal buoyancy and wind shear effects, which are captured by the sensible heat flux $\shfx$ and the friction velocity $u_*={\left(\overline{u'w'}^2 + \overline{v'w'}^2 \right)}^{1/4}$, respectively; where $\overline{u'w'}$ and $\overline{v'w'}$ are momentum flux components.
Additionally, we incorporate the variances of potential temperature and ho\-ri\-zon\-tal wind magnitude, $\sigma_\theta^2$ and $\sigma_M^2 = \sigma_u^2 + \sigma_v^2$.
The altitude $z$ serves as length scale because we aim for a surface layer $C_n^2$ parametrization, but 
to model OT at higher altitudes, the suitable length scale is expected to differ.
All relevant variables forming the input for our $\Pi$-ML methodology are summarized in the table of figure~\ref{fig:methodology}a with their respective fundamental dimensions.
Earth's gravitational acceleration $g=9.81 \text{m\,s}^{-2}$ is also included because it is required for the atmospheric force balance.
Given the dry atmospheric conditions at the MLO sites, moisture variables were ignored. 
To later assess the $C_n^2$ estimation performance of the trained model, the first two weeks of July 2006 are set aside as test data.
Although using data out of the middle as test data might seem unconventional, it is used so that the ML models can capture the seasonal change from June to August (see Appendix~\ref{app:testsets} for details).

The two key components of our proposed $\Pi$-ML methodology are illustrated in figure~\ref{fig:methodology}: 
the DA constrained with domain knowledge (a) and the ensemble of gradient-boosting ML models, which perform regression on the stacked, non-dimensionalized observations (b).
We set off with the table in figure~\ref{fig:methodology}a and the Buckingham $\Pi$ theorem \citep{stull1988}, popular in DA.
The theorem states that our $k=10$ dimensional variables with their $l=3$ fundamental dimensions (length, time, temperature) can be expressed as a set of $(k-l)=7$ independent non-dimensional $\Pi$ groups. 
Multiple options exist to form these sets, so we employ the $\Pi$ theorem implementation of \citep{karam2021}, which generates 71 sets with 7 $\Pi$ groups each.
Using domain knowledge, we conceive three constraints to reduce the number of sets from 71 to 14.
First, each set can only contain one dependent $\Pi$ group that is function of $C_n^2$ (cf.\ pink highlights in figure~\ref{fig:methodology}a). 
All other $\Pi$ groups should only be functions of the independent dimensional variables $\boldsymbol{X}$.
Second, $C_n^2$ and its normalized variant $\Pi_y$ vary over multiple orders of magnitude, so the ML models are trained on $\log_{10}\Pi_y$.
Since the logarithm is not defined for negative arguments, only $\Pi$ sets where $\Pi_y$ is strictly positive are valid.
Third, the dimensional variables $\Gamma$ and $\shfx$ can be positive and negative, so raising them to fractional or even-integer powers can result in complex values or a loss of sign.
Therefore, valid $\Pi$ sets cannot contain such expressions.
Each of the 14 constrained $\Pi$ sets is used to scale and non-dimensionalize the dimensional observations $\boldsymbol{X}$ and $y=C_n^2$ to yield $\boldsymbol{\Pi}_X$ and $\Pi_y$, respectively, as illustrated in figure~\ref{fig:methodology}b.
The non-dimensionalized observations from all three levels can be stacked into a combined dataset from which ML learns the non-dimensional black box similarity relationship $f(\boldsymbol{\Pi}_X) \approx \log_{10}\Pi_y$.
For each $\Pi$ set, we train one ensemble of $n=25$ member models to make robust $C_n^2$ predictions with uncertainty estimates using the gradient boosting algorithm XGBoost (XGB) and the AutoML library FLAML \citep{wang2021}.
FLAML performs time-constrained hyperparameter tuning of the XGB models using 5-fold cross-validation.
For each ensemble member, FLAML was given a 10-minute time budget on 8 cores of a 3\,GHz Intel Xeon E5-6248R CPU.
Such a time-constrained optimization is crucial to keep the overall training costs reasonable ($\sim$34 core hours per ensemble).
We employ the Monte-Carlo resampling strategy to generate a different 4-week subset of the 6-week training data for each member.
Two non-overlapping sets of seven consecutive days are randomly removed from the training data, so each subset covers  different meteorological conditions.
As depicted in figure~\ref{fig:methodology}b, each of the $n$ trained members produces a prediction that is robustly aggregated into an ensemble prediction using the median.
The prediction accuracy and model complexity of each trained $\Pi$-ML ensemble is assessed to decide which $\Pi$ set is best suited for our ML-based parameterization.
The root-mean-squared error (RMSE) $\epsilon = \sqrt{\langle{(y-\hat{y})}^2 \rangle}$ in the  log-space is utilized to quantify accuracy as the deviation between the observed $\log_{10}C_n^2=y$ from the test set (July 1 -- 14) and the corresponding $\Pi$-ML prediction $\hat{y}=\log_{10}\hat{C}_n^2$.
We also evaluate the complexity of the $\Pi$ sets and their trained ML ensembles. 
That is essential because ML models should only be as complex as necessary to increase their ability to perform well on new unseen data \citep{vapnik1998}.
One $\Pi$ set is considered simpler than another set if its $\Pi$ groups are constructed from fewer dimensional variables.
Similarly, one trained ensemble is considered simpler than another one if fewer $\Pi$ groups are important for the ML prediction, i.e.\ the modeled $C_n^2$ is sensitive to fewer input features.
The importance of input features of the trained $\Pi$-ML models is quantified by the permutation feature importance technique (PFI) \citep{molnar2022}.
For each feature $\Pi_i$, PFI yields a ratio $(\epsilon_i'-\epsilon)/\epsilon$, which describes how the RMSE $\epsilon_i$ of a trained model magnifies when the model gets shuffled data for $\Pi_i$ compared to the baseline RMSE $\epsilon$ where the correlation is intact.
That means a highly important feature results in a large error magnification.
\begin{figure}[t]
    \centering
    \begin{subfigure}[t]{0.53\columnwidth}
        \caption{}%
        \label{fig:scores_overview}
        \begin{center}
            \vspace{-6mm}
            \includegraphics{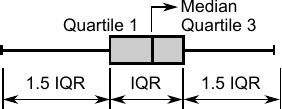}
            \includegraphics[width=\textwidth,trim={0mm 0mm 0mm -2mm}]{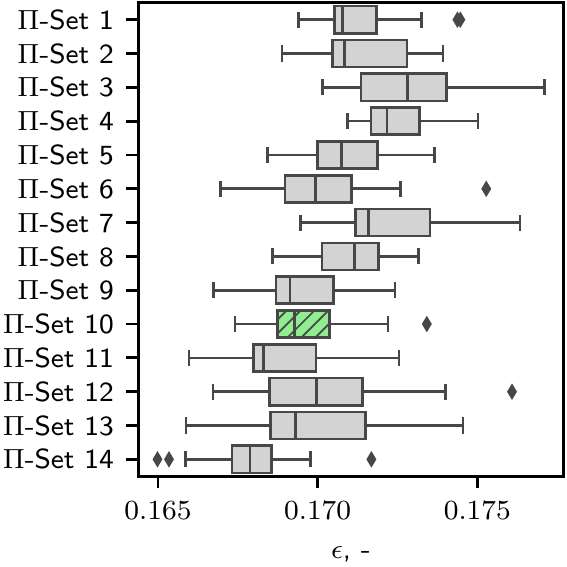}
        \end{center}
    \end{subfigure}\hfill
    \begin{subfigure}[t]{0.44\columnwidth}
        \caption{}%
        \label{fig:complexity_overview}
        \begin{center}
            \vspace{-3mm}
            \includegraphics[width=\textwidth,trim={6mm 3mm 2mm 9mm},clip]{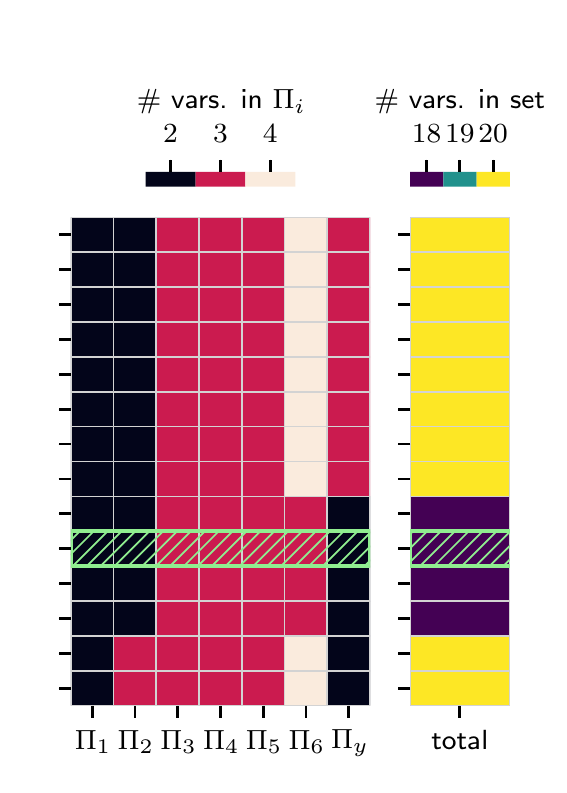}
        \end{center}
    \end{subfigure}
    \vspace{-2mm}
    \caption{Comparison of (a) ensemble performance and (b) $\Pi$ set complexity for our 14 different $\Pi$ sets, where winning set 10 (green/hatched) balances performance and complexity well.}%
    \label{fig:overview}
\end{figure}

\begin{figure}[t]
    \begin{center}
        \includegraphics[width=\columnwidth,trim={2mm 1mm 10mm 5mm},clip]{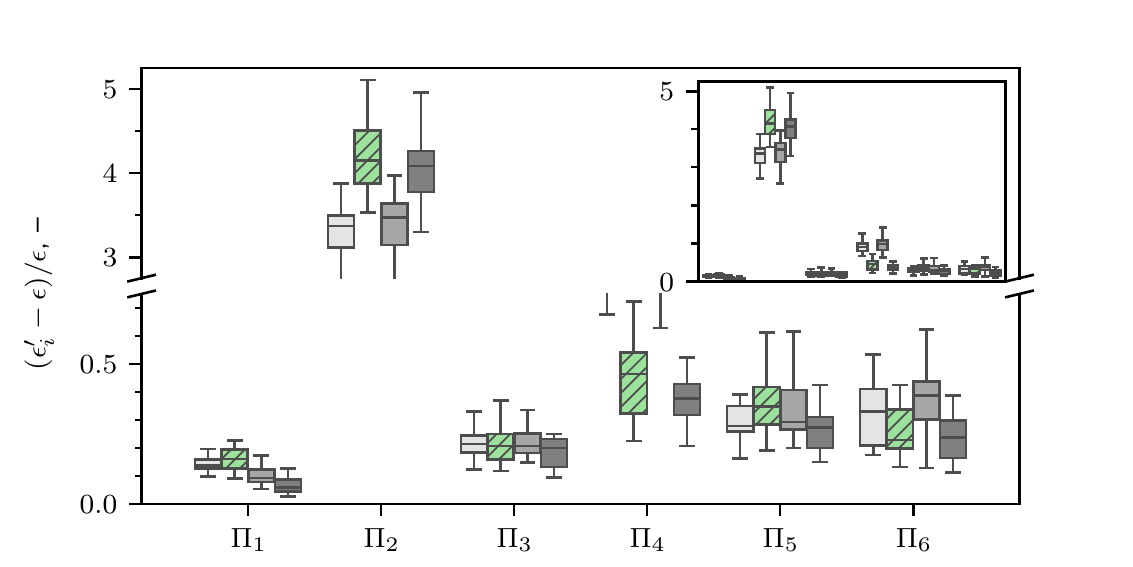}
    \end{center}
    \vspace{-4mm}
    \caption{Importance of non-dimensional $\Pi$ groups (i.e., $\Pi_1$ to $\Pi_6$) based on the permutation feature importance strategy. For easy intercomparison, for each non-dimensional $\Pi$-group, four boxplots representing ML ensembles corresponding to $\Pi$-sets 9, 10, 11, and 12 (left to right) are plotted side-by-side. The best performing ensemble 10 is marked in green (hatched).
    }%
    \label{fig:feature_importance}
\end{figure}

The performance and complexity of the 14 $\Pi$-ML ensembles are shown in figure~\ref{fig:overview}.
The boxplots in panel (a) display the $\epsilon$ distributions for each ensemble.
While all ensembles show median RMSEs of the same order of magnitude, $\Pi$ sets 9 to 14 outperform the others.
Panel (b) visualizes complexity through the number of dimensional variables constituting each $\Pi$ group (left) together with the sum per set (right). 
This plot reveals that sets 9 to 12 of the well-performing ensembles are the only ones consisting of $\Pi$ groups formed from no more than three dimensional variables.
These four low-error, low-complexity candidates are further assessed based on their PFI score distributions displayed in figure~\ref{fig:feature_importance}.
Remember that the DA yields different functional expressions for $\Pi_i$ for each set, which is why each set shows different PFI distributions.
The boxplots reveal that $\Pi$ sets 9 and 11 yield more complex $\Pi$-ML ensembles compared to 10 and 12 because they significantly rely on two $\Pi$ groups ($\Pi_2$ and $\Pi_4$, see inset) for $C_n^2$ estimation instead of one ($\Pi_2$).
Consequently, only sets 10 and 12 remain candidates for our ML-based similarity theory of optical turbulence.
From these, we ultimately select $\Pi$ set 10 because of the lower $\epsilon$ spread in figure~\ref{fig:scores_overview} with
$\Pi_1=\sigma_M^2/u_*^2$, $\Pi_2=\overline{\theta}/\sqrt{\sigma^2_\theta}$, $\Pi_3=(S\, z)/u_*$, $\Pi_4=\shfx/(u_* \sigma_\theta)$, $\Pi_5=(g\, z)/u_*^2$, $\Pi_6=(\Gamma\, z)/\sigma_\theta$, and $\Pi_y={(C_n^2)}^{3/2}\, z$.
The expressions for the other 13 $\Pi$ sets are listed in table~\ref{tab:full_set} of Appendix~\ref{app:full_set}.
The observation that $\Pi_2$ -- the inverse normalized potential temperature variance -- is the only dominating feature of our parameterization could have practical implications:
First, temperature variances can be measured with thermocouples \citep{albertson1995}, which are cheaper than sonic anemometers.
Second, the low relevance of the gradients ($\Pi_3$ and $\Pi_6$) indicates that even single-level measurements might be sufficient to estimate $C_n^2$ accurately.
Therefore, our approach might lead to simpler $C_n^2$ measurement setups. %
In Appendix~\ref{app:reduced}, we confirm that retraining the models with $\Pi_2$ as the sole input feature still yields highly accurate predictions.

\begin{figure}[t]
    \begin{center}
        \includegraphics[width=.95\columnwidth]{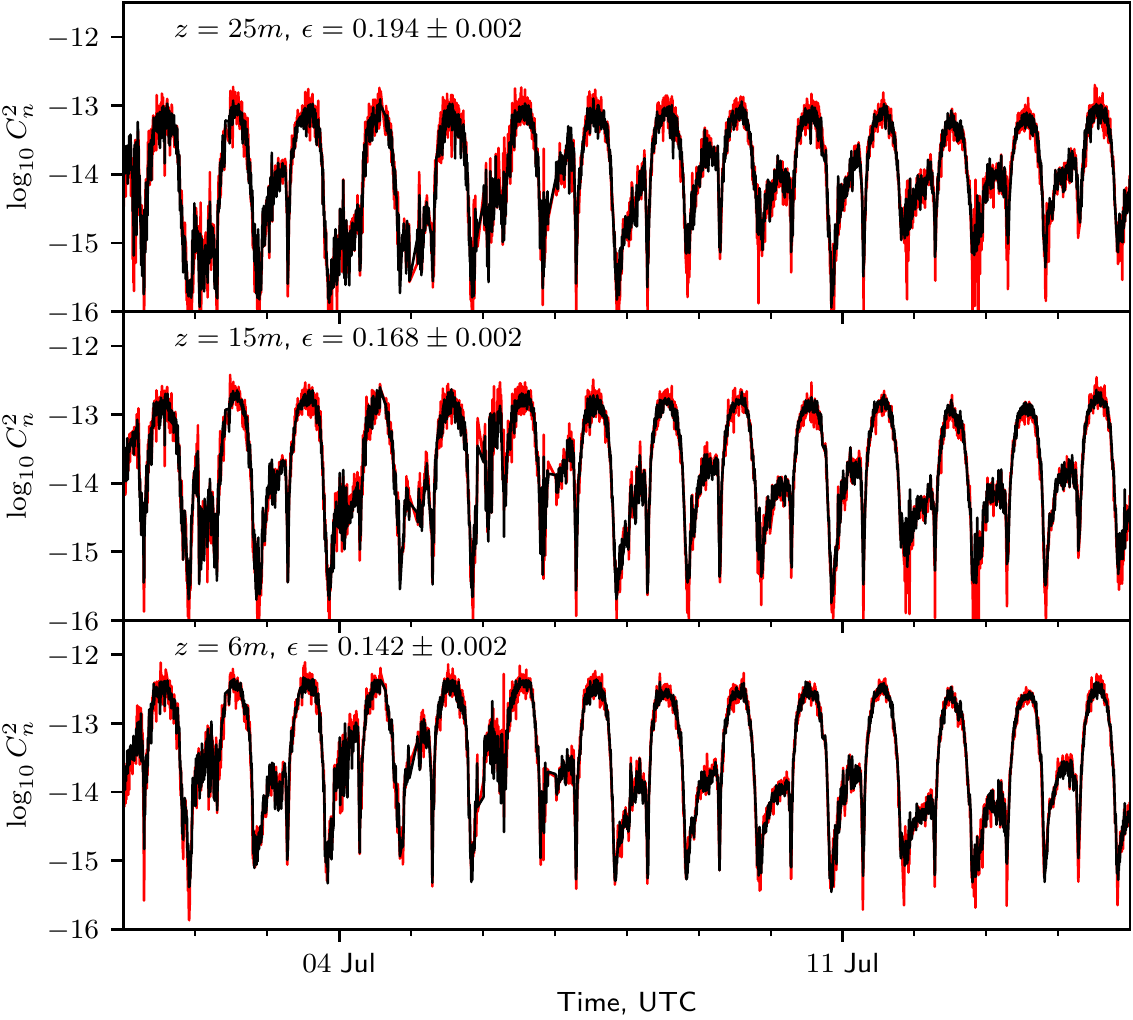}
    \end{center}
    \vspace{-5mm}
    \caption{
        Median predictions of $\log_{10}C_n^2$ based on test data (black) using the selected $\Pi$ set 10 ensemble. The observed values (red) are shown for reference. 
    }%
    \label{fig:prediction}
\end{figure}

\begin{figure}[t]
    \centering
    \begin{subfigure}[t]{0.54\columnwidth}
        \caption{}%
        \label{fig:correlation}
        \begin{center}
            \vspace{-6mm}
            \includegraphics[width=\textwidth]{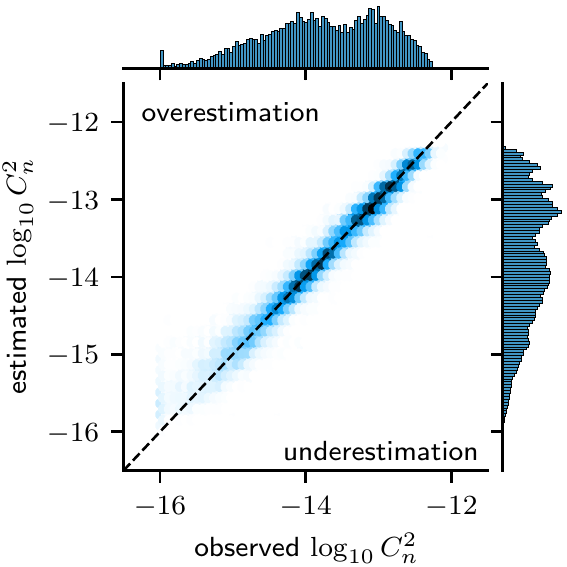}
        \end{center}
    \end{subfigure}\hfill
    \begin{subfigure}[t]{0.45\columnwidth}
        \caption{}%
        \label{fig:qqplot}
        \begin{center}
            \vspace{-6mm}
            \includegraphics[width=\textwidth,trim={5mm 0mm 4mm 1mm},clip]{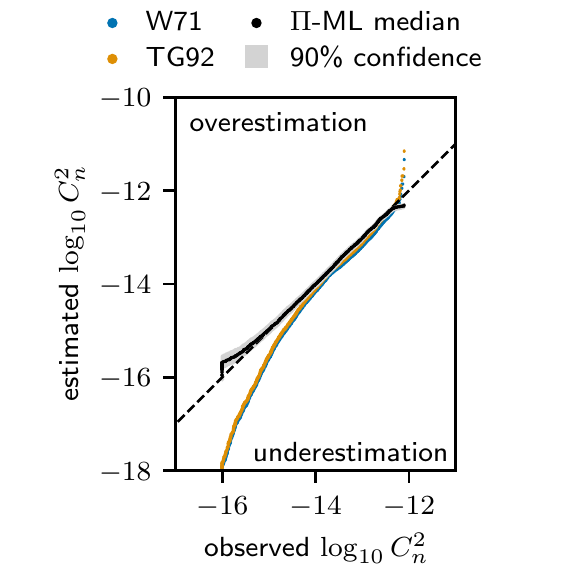}
        \end{center}
    \end{subfigure}
    \vspace{-2mm}
    \caption{
        Correlation histogram and quantile-quantile plot for $\Pi$ set 10 ensemble showing (a) high correlation ($R^2=0.958\pm0.001$) and (b) well-captured $C_n^2$ distributions compared to traditional models from literature (blue, orange).
    }%
    \label{fig:errors}
\end{figure}

The performance of the final $\Pi$-ML ensemble is illustrated in more detail in figures~\ref{fig:prediction} and \ref{fig:errors}.
The observed (red) and the predicted median evolutions of $C_n^2$ (black) for the test data are shown in figure~\ref{fig:prediction}.
The evolutions are plotted for the three original sonic heights individually for visualization.
The agreement between prediction and observation is high for all levels, although the level-specific $\epsilon$ slightly increases with height. 
For nighttime conditions, the surface layer depth is typically shallower than 10--20 m. Thus, the topmost sonic anemometer at 25 m might be outside the surface layer.
In addition, outer layer effects such as wave-induced bursting events can force the turbulence underneath \citep{mahrt2014}.
In such cases, cause (forcing) and effect (turbulence) are vertically separated, so the sonic signal only contains the effect but not the cause.
Thus, prediction accuracy decreases without additional upper-air information.
Notable errors on all levels mostly occur during atmospheric neutral conditions shortly after sunrise and sunset where the observed $C_n^2$ drops as low as $10^{-16}$.
These drops are overestimated by our ensemble, which is also visible in the 2D correlation histogram of figure~\ref{fig:correlation} and the quantile-quantile (QQ) plot in \ref{fig:qqplot}.
Panel (a) directly compares observed $C_n^2$ samples to their ML-estimated counterpart, 
while panel (b) plots the cumulative density functions of observed and estimated $C_n^2$ against each other.
The overestimation of neutral conditions is visible in both panels as the deviation of the histogram/curve from the ideal 1:1 line (dashed) for $C_n^2 < 10^{-15}$.
Simultaneously, the grey 90\% confidence band in (b) grows, which indicates increasing disagreement between the predictions of the ensemble members.
However, less than 8\% of $C_n^2$ measurements are smaller than $10^{-15}$, so the regularization of the ML training results in models that favor the center of the $C_n^2$ distribution, not its tails.
Also, the lower signal-to-noise ratio of the sonic anemometers in weak turbulence conditions increases the measurement uncertainty \citep{rannik2016}.
Since very low turbulence conditions are also not critical for optical applications such as optical links or astronomy, we argue that little emphasis should be put on these deviations. 
The regularization mentioned above also explains the minor underestimation visible in panel (b) for observations with $C_n^2 > 10^{-12.5}$, which make up less than 3.5\% of the data.
Leaving the tails of the distributions aside, both panels of figure~\ref{fig:errors} show excellent performance of our ensemble for most data.
Most points in (a) the histogram and (b) the QQ plot are close to the ideal 1:1 line as quantified by the coefficient of determination of $R^2=0.958$ computed on all test data, including the deviating tails. 
The spread of the correlation distribution around the 1:1 line is symmetric for $C_n^2 > 10^{-15}$.
That means the ensemble predictions are well-balanced and not biased towards over or underestimation for most of the $C_n^2$ range.
A brief comparison of $\Pi$-ML with two conventional MOST-based $C_n^2$ parameterizations (W71 \citep{wyngaard1971} and TG92 \citep{thiermann1992}) in figure~\ref{fig:qqplot} illustrates the potential of improvement by utilizing ML.
While W71 and TG92 have the operational advantage of being formulated as analytical equations, they lack the flexibility to capture complex behavior where ML excels.
This results in the larger over and underestimations shown in the QQ plots for these popular approaches.
Comparing $\Pi$-ML to a more traditional ML approach based on \cite{wang2016} (see Appendix~\ref{app:traditional}) also shows significantly higher performance of $\Pi$-ML.

In summary, we demonstrated how dimensional analysis constrained with domain knowledge yields non-dimensional surface layer scaling expressions, which enable us to train accurate XGBoost regression models.
Our approach has two advantages over $C_n^2$ parametrizations from the literature.
First, the final ensemble produced highly accurate predictions for both daytime and nighttime, while previous models are often limited to one or the other \citep{smith1993}.
Second, we expect that the non-dimensional formulation allows making predictions with a pre-trained ensemble for new sonics set up at different heights or locations 
if the new non-dimensionalized data fall into the original non-dimensional training ranges.
The data scaling should enable our ensemble to stay in the interpolation regime longer, i.e.\ cover a larger dimensional space, compared to traditional ML-based models. At this point, these claims are speculative in nature and need extensive validation.
Our final $\Pi$-ML ensemble was shown to perform well, regardless of the complex meteorology of Hawaii \citep{businger2002} and the limited measurement duration of only two months.
While the complexity and data sparsity of the MLO campaign limits the applicability of the trained ensemble to other sites, 
the good performance leads us to posit that our $\Pi$-ML methodology might perform well in more favorable setups.
Additionally, we observed a strong dependency of $C_n^2$ on $\sigma^2_\theta$ ($\Pi_2$), suggesting that relatively inexpensive single-level variance measurements might be sufficient for accurate $C_n^2$ estimation in the surface layer.
In conclusion, we presented a powerful, statistically robust physics-informed machine learning methodology ($\Pi$-ML) to estimate $C_n^2$ from turbulence measurements.
\begin{acknowledgements}
  MP is financed by the FREE project (P19-13) of the TTW-Perspectief research program partially financed by the Dutch Research Council (NWO).
We are grateful to NCAR for making the MLO $C_n^2$ data publicly available.
\end{acknowledgements}

\appendix

\section{Full list of non-dimensional $\Pi$ sets}%
\label{app:full_set}

\begin{table}[t]
  \centering
  \caption{Expressions of all non-dimensional $\Pi$ sets for input features ($\Pi_i$) and respective target variable $\Pi_y$ (scaled $C_n^2$). Note that $\sigma_X$ and $\sigma_X^2$ denote the standard deviation and variance of quantity $X$, respectively. The final set selected for its balance of ensemble prediction accuracy and complexity is marked with (*).}%
  \label{tab:full_set}
  \vspace{6pt}
  \renewcommand{\arraystretch}{1.75}
  \begin{tabular}{lccccccc}
    \toprule
    Set & $\Pi_1$ & $\Pi_2$ & $\Pi_3$ & $\Pi_4$ & $\Pi_5$ & $\Pi_6$ & $\Pi_y$  \\
    \midrule
    Set 1 & $\frac{\sigma_M^2}{u_*^{2}}$ & $\frac{\sigma_\theta^2}{\overline{\theta}^{2}}$ & $\frac{S\,z}{u_*}$ & $\frac{\overline{w'\theta'}}{\overline{\theta}\,u_*}$ & $\frac{g}{S\,u_*}$ & $\frac{\Gamma\,u_*}{S\,\overline{\theta}}$ & $\frac{u_*\,(C_n^2)^{3/2}}{S}$\\
Set 2 & $\frac{\sigma_M^2}{u_*^{2}}$ & $\frac{\overline{\theta}}{\sigma_\theta}$ & $\frac{S\,z}{u_*}$ & $\frac{\overline{w'\theta'}}{u_*\,\sigma_\theta}$ & $\frac{g}{S\,u_*}$ & $\frac{\Gamma\,u_*}{S\,\sigma_\theta}$ & $\frac{u_*\,(C_n^2)^{3/2}}{S}$\\
Set 3 & $\frac{\sigma_M^2}{u_*^{2}}$ & $\frac{\sigma_\theta^2}{\overline{\theta}^{2}}$ & $\frac{g\,z}{u_*^{2}}$ & $\frac{S\,u_*}{g}$ & $\frac{\overline{w'\theta'}}{\overline{\theta}\,u_*}$ & $\frac{\Gamma\,u_*^{2}}{g\,\overline{\theta}}$ & $\frac{u_*^{2}\,(C_n^2)^{3/2}}{g}$\\
Set 4 & $\frac{\sigma_M^2}{u_*^{2}}$ & $\frac{\overline{\theta}}{\sigma_\theta}$ & $\frac{g\,z}{u_*^{2}}$ & $\frac{S\,u_*}{g}$ & $\frac{\overline{w'\theta'}}{u_*\,\sigma_\theta}$ & $\frac{\Gamma\,u_*^{2}}{g\,\sigma_\theta}$ & $\frac{u_*^{2}\,(C_n^2)^{3/2}}{g}$\\
Set 5 & $\frac{u_*}{\sigma_M}$ & $\frac{\sigma_\theta^2}{\overline{\theta}^{2}}$ & $\frac{S\,z}{\sigma_M}$ & $\frac{\overline{w'\theta'}}{\overline{\theta}\,\sigma_M}$ & $\frac{g}{S\,\sigma_M}$ & $\frac{\Gamma\,\sigma_M}{S\,\overline{\theta}}$ & $\frac{\sigma_M\,(C_n^2)^{3/2}}{S}$\\
Set 6 & $\frac{u_*}{\sigma_M}$ & $\frac{\overline{\theta}}{\sigma_\theta}$ & $\frac{S\,z}{\sigma_M}$ & $\frac{\overline{w'\theta'}}{\sigma_\theta\,\sigma_M}$ & $\frac{g}{S\,\sigma_M}$ & $\frac{\Gamma\,\sigma_M}{S\,\sigma_\theta}$ & $\frac{\sigma_M\,(C_n^2)^{3/2}}{S}$\\
Set 7 & $\frac{u_*}{\sigma_M}$ & $\frac{\sigma_\theta^2}{\overline{\theta}^{2}}$ & $\frac{g\,z}{\sigma_M^2}$ & $\frac{S\,\sigma_M}{g}$ & $\frac{\overline{w'\theta'}}{\overline{\theta}\,\sigma_M}$ & $\frac{\Gamma\,\sigma_M^2}{g\,\overline{\theta}}$ & $\frac{\sigma_M^2\,(C_n^2)^{3/2}}{g}$\\
Set 8 & $\frac{u_*}{\sigma_M}$ & $\frac{\overline{\theta}}{\sigma_\theta}$ & $\frac{g\,z}{\sigma_M^2}$ & $\frac{S\,\sigma_M}{g}$ & $\frac{\overline{w'\theta'}}{\sigma_\theta\,\sigma_M}$ & $\frac{\Gamma\,\sigma_M^2}{g\,\sigma_\theta}$ & $\frac{\sigma_M^2\,(C_n^2)^{3/2}}{g}$\\
Set 9 & $\frac{\sigma_M^2}{u_*^{2}}$ & $\frac{\sigma_\theta^2}{\overline{\theta}^{2}}$ & $\frac{S\,z}{u_*}$ & $\frac{\overline{w'\theta'}}{\overline{\theta}\,u_*}$ & $\frac{g\,z}{u_*^{2}}$ & $\frac{\Gamma\,z}{\overline{\theta}}$ & $(C_n^2)^{3/2}\,z$\\
Set 10* & $\frac{\sigma_M^2}{u_*^{2}}$ & $\frac{\overline{\theta}}{\sigma_\theta}$ & $\frac{S\,z}{u_*}$ & $\frac{\overline{w'\theta'}}{u_*\,\sigma_\theta}$ & $\frac{g\,z}{u_*^{2}}$ & $\frac{\Gamma\,z}{\sigma_\theta}$ & $(C_n^2)^{3/2}\,z$\\
Set 11 & $\frac{u_*}{\sigma_M}$ & $\frac{\sigma_\theta^2}{\overline{\theta}^{2}}$ & $\frac{S\,z}{\sigma_M}$ & $\frac{\overline{w'\theta'}}{\overline{\theta}\,\sigma_M}$ & $\frac{g\,z}{\sigma_M^2}$ & $\frac{\Gamma\,z}{\overline{\theta}}$ & $(C_n^2)^{3/2}\,z$\\
Set 12 & $\frac{u_*}{\sigma_M}$ & $\frac{\overline{\theta}}{\sigma_\theta}$ & $\frac{S\,z}{\sigma_M}$ & $\frac{\overline{w'\theta'}}{\sigma_\theta\,\sigma_M}$ & $\frac{g\,z}{\sigma_M^2}$ & $\frac{\Gamma\,z}{\sigma_\theta}$ & $(C_n^2)^{3/2}\,z$\\
Set 13 & $\frac{\sigma_\theta^2}{\overline{\theta}^{2}}$ & $\frac{u_*}{\sqrt{g}\,\sqrt{z}}$ & $\frac{S\,\sqrt{z}}{\sqrt{g}}$ & $\frac{\sigma_M^2}{g\,z}$ & $\frac{\Gamma\,z}{\overline{\theta}}$ & $\frac{\overline{w'\theta'}}{\sqrt{g}\,\overline{\theta}\,\sqrt{z}}$ & $(C_n^2)^{3/2}\,z$\\
Set 14 & $\frac{\overline{\theta}}{\sigma_\theta}$ & $\frac{u_*}{\sqrt{g}\,\sqrt{z}}$ & $\frac{S\,\sqrt{z}}{\sqrt{g}}$ & $\frac{\sigma_M^2}{g\,z}$ & $\frac{\Gamma\,z}{\sigma_\theta}$ & $\frac{\overline{w'\theta'}}{\sqrt{g}\,\sigma_\theta\,\sqrt{z}}$ & $(C_n^2)^{3/2}\,z$\\

    \bottomrule
  \end{tabular}
\end{table}

Table~\ref{tab:full_set} supplements figure~\ref{fig:overview} and provides a full list of the non-dimensional expressions of the 14 $\Pi$ sets.
The $\Pi_i$ groups are sorted by complexity just like in figure~\ref{fig:complexity_overview}, and the normalized $C_n^2$ target is denoted by $\Pi_y$.

\section{Performance evaluation using different test set intervals}%
\label{app:testsets}

\begin{figure*}
  \centering
  \begin{subfigure}[t]{.32\textwidth}
    \caption{Test set interval (*):\\ 2006-06-09 -- 2006-06-29}
    \vspace{1.05em}
    \includegraphics[width=\textwidth]{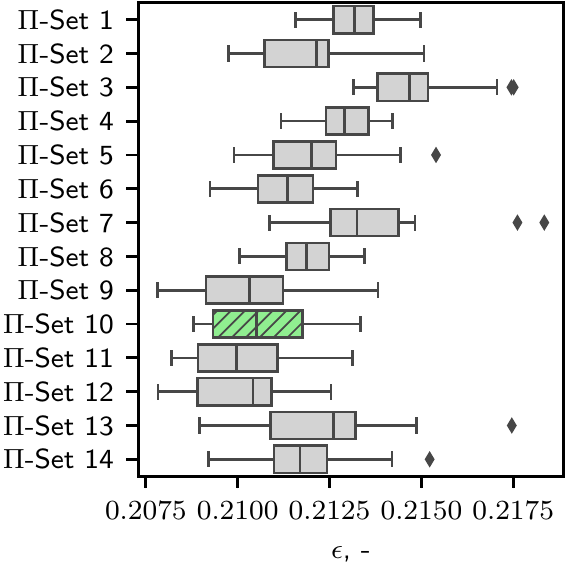}
  \end{subfigure}\hfill
  \begin{subfigure}[t]{.32\textwidth}
    \caption{Test set interval:\\ 2006-07-01 -- 2006-07-15\\ (repeated from main text)}
    \includegraphics[width=\textwidth]{figures/02a_score_overview.pdf}
  \end{subfigure}\hfill
  \begin{subfigure}[t]{.32\textwidth}
    \caption{Test set interval:\\ 2006-07-25 -- 2006-08-09}
    \vspace{1.05em}
    \includegraphics[width=\textwidth]{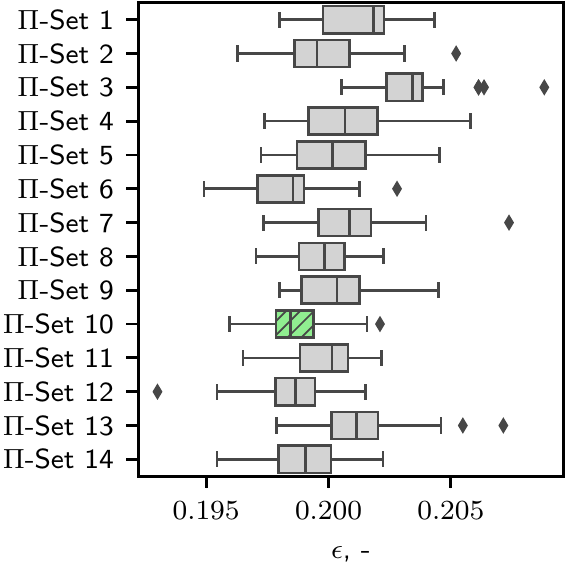}
  \end{subfigure}
  \caption{Comparison of model performance when three different test set intervals are used. 
  Each interval contains 14 days of data for testing. 
  The first interval marked with (*) contains a few days of missing data, so the window is large enough to contain 14 full valid days.}%
  \label{fig:testsets}
\end{figure*}

In the main text, a 14-day test set (July 1-14, 2006) was used to evaluate all the $\Pi$-ML and conventional models. This test set was extracted from the middle of the MLO study dataset; the rest of the dataset was utilized in training. This train-test split strategy enables the models to `learn' seasonal trends in local weather patterns. Instead, if the test data were taken from the beginning or toward the end of the dataset, the trained $\Pi$-ML would need to extrapolate the trends rather than interpolate them. In figure~\ref{fig:testsets}, we report the sensitivity of the model performance with respect to different test set intervals. Subplots (a) and (c) represent the ensemble performance if test data is taken from the beginning or toward the end of the dataset, respectively. Subplot (b) is a repetition of the boxplot from figure~\ref{fig:overview} of the main text for completeness. These plots reveal that the absolute root-mean-squared-error $\epsilon$ increases by $\sim 20\%$ for cases (a) and (c) from corresponding values in case (b). Similarly, the coefficient of determination $R^2$ of the selected $\Pi$-set 10 drops marginally from ca. $0.96$ in case (b) to ca. $0.93$ in cases (a) and (c). These degradations in performance are expected due to extrapolation beyond the training range. %

\section{Performance of a $\Pi$-ML model of reduced complexity}%
\label{app:reduced}

\begin{figure*}
  \centering
  \begin{minipage}[c]{.55\textwidth}
      \begin{subfigure}[t]{\textwidth}
        \caption{Median predictions (black) compared to test set data (red) per $z$-level.}
        \includegraphics[width=\textwidth]{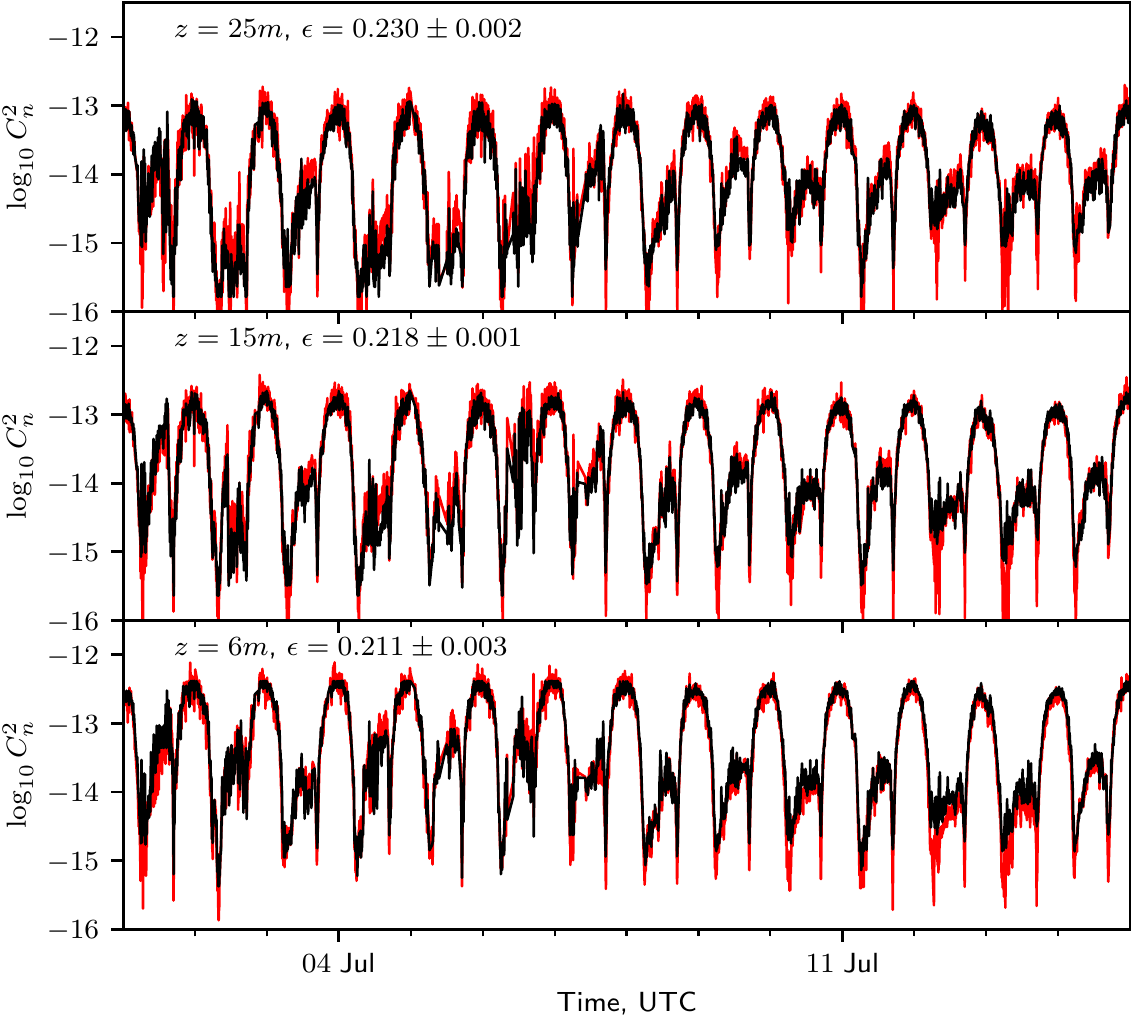}
      \end{subfigure}
  \end{minipage}
  \hspace{2em}
  \begin{minipage}[c]{.3\textwidth}
    \begin{subfigure}[t]{\textwidth}
      \caption{Correlation histogram with\\ $R^2=0.929\pm0.000$}
      \includegraphics[width=\textwidth]{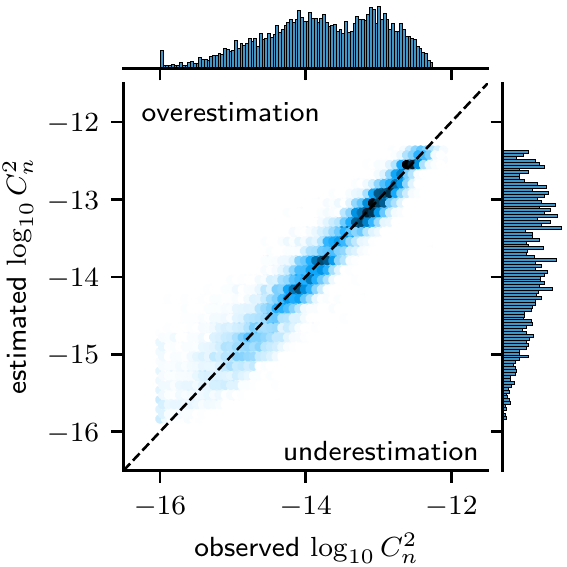}
    \end{subfigure}

    \begin{subfigure}[t]{\textwidth}
      \caption{QQ plot}
      \includegraphics[width=\textwidth]{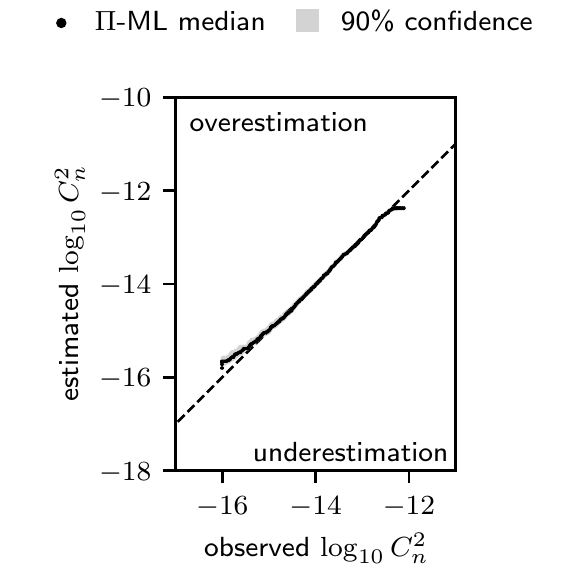}
    \end{subfigure}
  \end{minipage}
  \caption{Performance of $\Pi$-ML model trained with $\Pi_2=\overline{\theta}/\sigma_\theta$ as sole feature and $\Pi_y = {(C_n^2)}^{3/2}\,z$ as target. Refer to figures 4 and 5 of the main text where the model performance of the full $\Pi$-set 10 model is displayed in equivalent formats.}%
  \label{fig:red13}
\end{figure*}

\begin{figure*}[t]
  \centering
  \includegraphics[width=.9\textwidth]{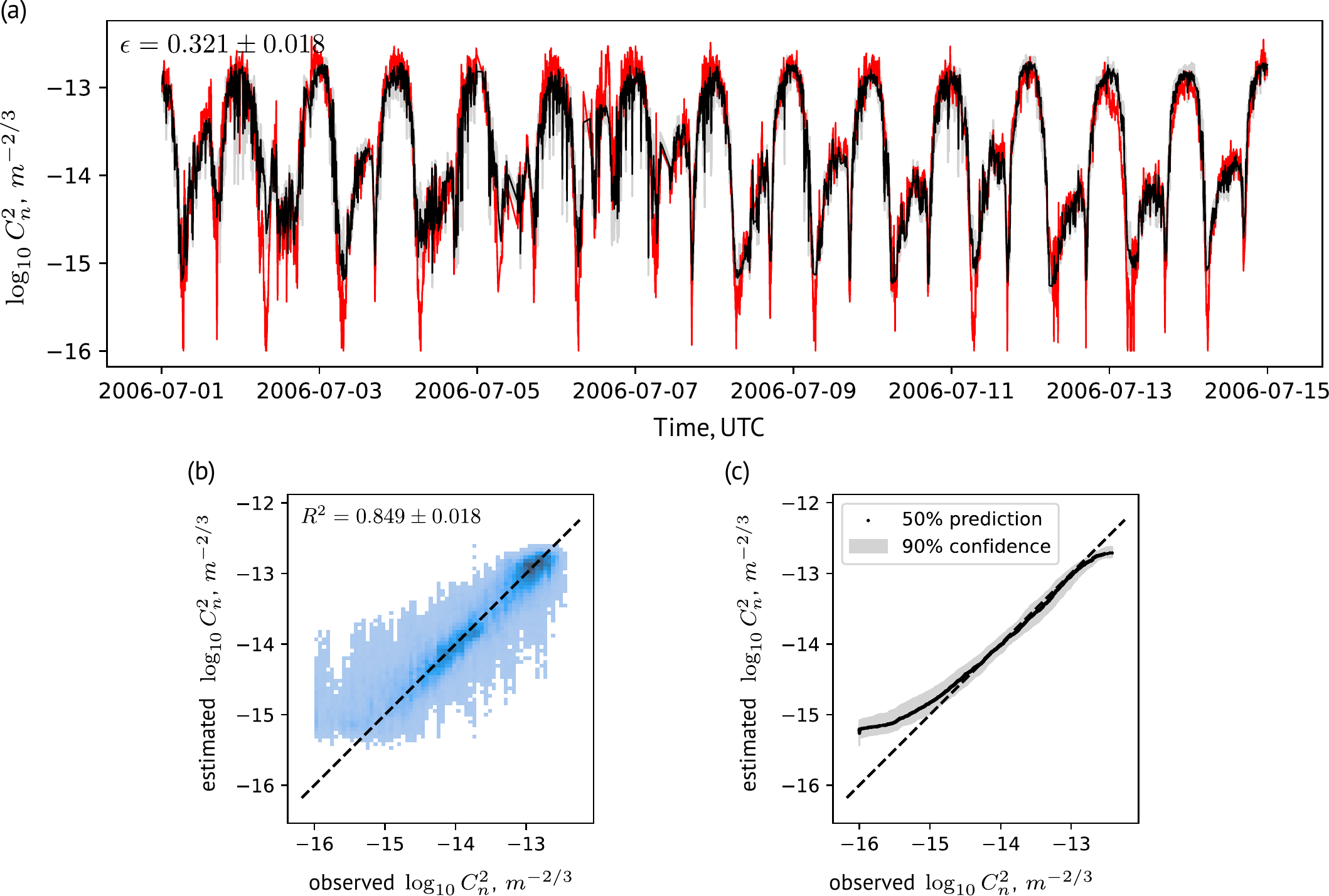}
  \caption{XGBoost ensemble trained on traditional features (2m-temperature, 2m-relative humidity, 2m-pressure, 15m-$\Gamma$, 15m-$S$) as proposed by Wang and Basu (2016) [7] to predict 15m-$C_n^2$. The predicted (black) and observed (red) 15m time series are shown in (a) with correlation plot and quantile-quantile plot in panels (b) and (c).}%
  \label{fig:wb16_comparison}
\end{figure*}

Based on feature importance analysis, shown in figure~\ref{fig:feature_importance} of the main text, we concluded that   $\Pi_2=\overline{\theta}/\sigma_\theta$ is the most dominant input feature predicting $C_n^2$ scaled as $\Pi_y = {(C_n^2)}^{3/2}\,z$.
To further support this claim, we trained a new reduced-order $\Pi$-ML model with only $\Pi_2$ as input.
The results are displayed in figure~\ref{fig:red13}. These plots should be compared against figures~\ref{fig:prediction} and \ref{fig:errors} of the main text. 
The performance of the new reduced-order model deteriorates slightly compared to the baseline model, which includes all the $\Pi$ groups from $\Pi$-set 10. The mean RMSE increases from $\epsilon=0.170\pm0.001$ (baseline model) to $\epsilon=0.220\pm0.000$ (reduced-order model), and mean correlation decreases from $R^2=0.958\pm0.001$ (baseline model) to $R^2=0.929\pm0.000$ (reduced-order model). Such drops in performance are expected because the contribution of $\Pi_i$ other than $\Pi_2$, in the baseline model is small but not negligible. Nonetheless, the overall performance of the new, reduced-order $\Pi$-ML model is very satisfactory; it is surprising that the XGBoost ensemble can predict surface layer $C_n^2$ values from just $\sigma_\theta$ and $\overline{\theta}$ as inputs.
We would like to emphasize that this result does not mean that the other $\Pi$ groups carry no physical importance or meaning. 
It only demonstrates that $\Pi_2$ alone already contains enough information for our $\Pi$-ML model to accurately predict $C_n^2$.

\section{Performance of an XGBoost ensemble trained with traditional features}%
\label{app:traditional}

To corroborate our claim that non-dimensional features are better suited for $C_n^2$ prediction than dimensional ones, we trained an XGBoost ensemble based on traditional features using our XGBoost ensemble strategy.
The traditional dimensional inputs are 2m-temperature, 2m-relative humidity, 2m-pressure, 15m-$\Gamma$, and 15m-$S$ to predict the 15m-$C_n^2$ as proposed by Wang and Basu (2016) [7].
The performance of the resulting model is shown in figure~\ref{fig:wb16_comparison}. 
Instead of the multilayer perceptron used by the original authors, we use our XGBoost ensemble methodology, which we also use for $\Pi$-ML. 
Although this approach leads to a reasonable result, the root-mean-square error $\epsilon$ using the traditional features is almost twice as high as for our best $\Pi$-ML ensemble, and $R^2$ is also notably lower. 
Additionally, the non-dimensional scaling enables $\Pi$-ML to predict $C_n^2$ independent of the $z$ level, while the traditional approach presented in figure~\ref{fig:wb16_comparison} requires to train one model per $z$ level.

\bibliography{literature_long.bib}%

\begin{thebibliography}{22}%
\makeatletter
\providecommand \@ifxundefined [1]{%
 \@ifx{#1\undefined}
}%
\providecommand \@ifnum [1]{%
 \ifnum #1\expandafter \@firstoftwo
 \else \expandafter \@secondoftwo
 \fi
}%
\providecommand \@ifx [1]{%
 \ifx #1\expandafter \@firstoftwo
 \else \expandafter \@secondoftwo
 \fi
}%
\providecommand \natexlab [1]{#1}%
\providecommand \enquote  [1]{``#1''}%
\providecommand \bibnamefont  [1]{#1}%
\providecommand \bibfnamefont [1]{#1}%
\providecommand \citenamefont [1]{#1}%
\providecommand \href@noop [0]{\@secondoftwo}%
\providecommand \href [0]{\begingroup \@sanitize@url \@href}%
\providecommand \@href[1]{\@@startlink{#1}\@@href}%
\providecommand \@@href[1]{\endgroup#1\@@endlink}%
\providecommand \@sanitize@url [0]{\catcode `\\12\catcode `\$12\catcode
  `\&12\catcode `\#12\catcode `\^12\catcode `\_12\catcode `\%12\relax}%
\providecommand \@@startlink[1]{}%
\providecommand \@@endlink[0]{}%
\providecommand \url  [0]{\begingroup\@sanitize@url \@url }%
\providecommand \@url [1]{\endgroup\@href {#1}{\urlprefix }}%
\providecommand \urlprefix  [0]{URL }%
\providecommand \Eprint [0]{\href }%
\providecommand \doibase [0]{https://doi.org/}%
\providecommand \selectlanguage [0]{\@gobble}%
\providecommand \bibinfo  [0]{\@secondoftwo}%
\providecommand \bibfield  [0]{\@secondoftwo}%
\providecommand \translation [1]{[#1]}%
\providecommand \BibitemOpen [0]{}%
\providecommand \bibitemStop [0]{}%
\providecommand \bibitemNoStop [0]{.\EOS\space}%
\providecommand \EOS [0]{\spacefactor3000\relax}%
\providecommand \BibitemShut  [1]{\csname bibitem#1\endcsname}%
\let\auto@bib@innerbib\@empty
\bibitem [{\citenamefont {Hemmati}(2009)}]{Hemmati2009}%
  \BibitemOpen
  \bibfield  {author} {\bibinfo {author} {\bibfnamefont {H.}~\bibnamefont
  {Hemmati}},\ }\href@noop {} {\emph {\bibinfo {title} {Near-Earth Laser
  Communications}}}\ (\bibinfo  {publisher} {{CRC Press}},\ \bibinfo {year}
  {2009})\BibitemShut {NoStop}%
\bibitem [{\citenamefont {Kaushal}\ and\ \citenamefont
  {Kaddoum}(2017)}]{kaushal2017}%
  \BibitemOpen
  \bibfield  {author} {\bibinfo {author} {\bibfnamefont {H.}~\bibnamefont
  {Kaushal}}\ and\ \bibinfo {author} {\bibfnamefont {G.}~\bibnamefont
  {Kaddoum}},\ }\bibfield  {title} {\bibinfo {title} {Optical {{Communication}}
  in {{Space}}: {{Challenges}} and {{Mitigation Techniques}}},\ }\href
  {https://doi.org/10.1109/COMST.2016.2603518} {\bibfield  {journal} {\bibinfo
  {journal} {IEEE Communications Surveys \& Tutorials}\ }\textbf {\bibinfo
  {volume} {19}},\ \bibinfo {pages} {57} (\bibinfo {year} {2017})},\ \Eprint
  {https://arxiv.org/abs/1705.10630} {arxiv:1705.10630 [cs, math]} \BibitemShut
  {NoStop}%
\bibitem [{\citenamefont {Wyngaard}\ \emph {et~al.}(1971)\citenamefont
  {Wyngaard}, \citenamefont {Izumi},\ and\ \citenamefont
  {Collins}}]{wyngaard1971}%
  \BibitemOpen
  \bibfield  {author} {\bibinfo {author} {\bibfnamefont {J.~C.}\ \bibnamefont
  {Wyngaard}}, \bibinfo {author} {\bibfnamefont {Y.}~\bibnamefont {Izumi}},\
  and\ \bibinfo {author} {\bibfnamefont {S.~A.}\ \bibnamefont {Collins}},\
  }\bibfield  {title} {\bibinfo {title} {Behavior of the
  {{Refractive-Index-Structure Parameter}} near the {{Ground}}*},\ }\href
  {https://doi.org/10.1364/JOSA.61.001646} {\bibfield  {journal} {\bibinfo
  {journal} {Journal of the Optical Society of America}\ }\textbf {\bibinfo
  {volume} {61}},\ \bibinfo {pages} {1646} (\bibinfo {year}
  {1971})}\BibitemShut {NoStop}%
\bibitem [{\citenamefont {Smith}\ \emph {et~al.}(1993)\citenamefont {Smith},
  \citenamefont {Accetta},\ and\ \citenamefont {Shumaker}}]{smith1993}%
  \BibitemOpen
  \bibfield  {author} {\bibinfo {author} {\bibfnamefont {F.~G.}\ \bibnamefont
  {Smith}}, \bibinfo {author} {\bibfnamefont {J.~S.}\ \bibnamefont {Accetta}},\
  and\ \bibinfo {author} {\bibfnamefont {D.~L.}\ \bibnamefont {Shumaker}},\
  }\href@noop {} {\emph {\bibinfo {title} {Atmospheric {{Propagation}} of
  {{Radiation}}}}},\ \bibinfo {series} {The {{Infrared}} \& {{Electro-Optical
  Systems Handbook}}}, Vol.~\bibinfo {volume} {2}\ (\bibinfo  {publisher}
  {{Infrared Information Analysis Center}},\ \bibinfo {year}
  {1993})\BibitemShut {NoStop}%
\bibitem [{\citenamefont {Monin}\ and\ \citenamefont
  {Obukhov}(1954)}]{monin1954}%
  \BibitemOpen
  \bibfield  {author} {\bibinfo {author} {\bibfnamefont {A.~S.}\ \bibnamefont
  {Monin}}\ and\ \bibinfo {author} {\bibfnamefont {A.~M.}\ \bibnamefont
  {Obukhov}},\ }\bibfield  {title} {\bibinfo {title} {Basic laws of turbulent
  mixing in the surface layer of the atmosphere},\ }\href@noop {} {\bibfield
  {journal} {\bibinfo  {journal} {Contrib. Geophys. Inst. Acad. Sci.}\ }\textbf
  {\bibinfo {volume} {151}} (\bibinfo {year} {1954})}\BibitemShut {NoStop}%
\bibitem [{\citenamefont {Savage}(2009)}]{savage2009}%
  \BibitemOpen
  \bibfield  {author} {\bibinfo {author} {\bibfnamefont {M.~J.}\ \bibnamefont
  {Savage}},\ }\bibfield  {title} {\bibinfo {title} {Estimation of evaporation
  using a dual-beam surface layer scintillometer and component energy balance
  measurements},\ }\href {https://doi.org/10.1016/j.agrformet.2008.09.012}
  {\bibfield  {journal} {\bibinfo  {journal} {Agricultural and Forest
  Meteorology}\ }\textbf {\bibinfo {volume} {149}},\ \bibinfo {pages} {501}
  (\bibinfo {year} {2009})}\BibitemShut {NoStop}%
\bibitem [{\citenamefont {Wang}\ and\ \citenamefont {Basu}(2016)}]{wang2016}%
  \BibitemOpen
  \bibfield  {author} {\bibinfo {author} {\bibfnamefont {Y.}~\bibnamefont
  {Wang}}\ and\ \bibinfo {author} {\bibfnamefont {S.}~\bibnamefont {Basu}},\
  }\bibfield  {title} {\bibinfo {title} {Using an artificial neural network
  approach to estimate surface-layer optical turbulence at {{Mauna Loa}},
  {{Hawaii}}},\ }\href {https://doi.org/10.1364/OL.41.002334} {\bibfield
  {journal} {\bibinfo  {journal} {Optics Letters}\ }\textbf {\bibinfo {volume}
  {41}},\ \bibinfo {pages} {2334} (\bibinfo {year} {2016})}\BibitemShut
  {NoStop}%
\bibitem [{\citenamefont {Jellen}\ \emph {et~al.}(2021)\citenamefont {Jellen},
  \citenamefont {Oakley}, \citenamefont {Nelson}, \citenamefont {Burkhardt},\
  and\ \citenamefont {Brownell}}]{jellen2021}%
  \BibitemOpen
  \bibfield  {author} {\bibinfo {author} {\bibfnamefont {C.}~\bibnamefont
  {Jellen}}, \bibinfo {author} {\bibfnamefont {M.}~\bibnamefont {Oakley}},
  \bibinfo {author} {\bibfnamefont {C.}~\bibnamefont {Nelson}}, \bibinfo
  {author} {\bibfnamefont {J.}~\bibnamefont {Burkhardt}},\ and\ \bibinfo
  {author} {\bibfnamefont {C.}~\bibnamefont {Brownell}},\ }\bibfield  {title}
  {\bibinfo {title} {Machine-learning informed macro-meteorological models for
  the near-maritime environment},\ }\href {https://doi.org/10.1364/AO.416680}
  {\bibfield  {journal} {\bibinfo  {journal} {Applied Optics}\ }\textbf
  {\bibinfo {volume} {60}},\ \bibinfo {pages} {2938} (\bibinfo {year}
  {2021})}\BibitemShut {NoStop}%
\bibitem [{\citenamefont {Bolbasova}\ \emph {et~al.}(2021)\citenamefont
  {Bolbasova}, \citenamefont {Andrakhanov},\ and\ \citenamefont
  {Shikhovtsev}}]{bolbasova2021}%
  \BibitemOpen
  \bibfield  {author} {\bibinfo {author} {\bibfnamefont {L.~A.}\ \bibnamefont
  {Bolbasova}}, \bibinfo {author} {\bibfnamefont {A.~A.}\ \bibnamefont
  {Andrakhanov}},\ and\ \bibinfo {author} {\bibfnamefont {A.~Y.}\ \bibnamefont
  {Shikhovtsev}},\ }\bibfield  {title} {\bibinfo {title} {The application of
  machine learning to predictions of optical turbulence in the surface layer at
  {{Baikal Astrophysical Observatory}}},\ }\href
  {https://doi.org/10.1093/mnras/stab953} {\bibfield  {journal} {\bibinfo
  {journal} {Monthly Notices of the Royal Astronomical Society}\ }\textbf
  {\bibinfo {volume} {504}},\ \bibinfo {pages} {6008} (\bibinfo {year}
  {2021})}\BibitemShut {NoStop}%
\bibitem [{\citenamefont {Su}\ \emph {et~al.}(2021)\citenamefont {Su},
  \citenamefont {Wu}, \citenamefont {Wu}, \citenamefont {Yang}, \citenamefont
  {Han}, \citenamefont {Qing}, \citenamefont {Luo},\ and\ \citenamefont
  {Liu}}]{su2021}%
  \BibitemOpen
  \bibfield  {author} {\bibinfo {author} {\bibfnamefont {C.}~\bibnamefont
  {Su}}, \bibinfo {author} {\bibfnamefont {X.}~\bibnamefont {Wu}}, \bibinfo
  {author} {\bibfnamefont {S.}~\bibnamefont {Wu}}, \bibinfo {author}
  {\bibfnamefont {Q.}~\bibnamefont {Yang}}, \bibinfo {author} {\bibfnamefont
  {Y.}~\bibnamefont {Han}}, \bibinfo {author} {\bibfnamefont {C.}~\bibnamefont
  {Qing}}, \bibinfo {author} {\bibfnamefont {T.}~\bibnamefont {Luo}},\ and\
  \bibinfo {author} {\bibfnamefont {Y.}~\bibnamefont {Liu}},\ }\bibfield
  {title} {\bibinfo {title} {In situ measurements and neural network analysis
  of the profiles of optical turbulence over the {{Tibetan Plateau}}},\ }\href
  {https://doi.org/10.1093/mnras/stab1792} {\bibfield  {journal} {\bibinfo
  {journal} {Monthly Notices of the Royal Astronomical Society}\ }\textbf
  {\bibinfo {volume} {506}},\ \bibinfo {pages} {3430} (\bibinfo {year}
  {2021})}\BibitemShut {NoStop}%
\bibitem [{\citenamefont {Stull}(1988)}]{stull1988}%
  \BibitemOpen
  \bibfield  {author} {\bibinfo {author} {\bibfnamefont {R.~B.}\ \bibnamefont
  {Stull}},\ }\href@noop {} {\emph {\bibinfo {title} {An Introduction to
  Boundary Layer Meteorology}}}\ (\bibinfo  {publisher} {{Kluwer Academic
  Publishers}},\ \bibinfo {address} {{Dordrecht}},\ \bibinfo {year}
  {1988})\BibitemShut {NoStop}%
\bibitem [{\citenamefont {Kashinath}\ \emph {et~al.}(2021)\citenamefont
  {Kashinath}, \citenamefont {Mustafa}, \citenamefont {Albert}, \citenamefont
  {Wu}, \citenamefont {Jiang}, \citenamefont {Esmaeilzadeh}, \citenamefont
  {Azizzadenesheli}, \citenamefont {Wang}, \citenamefont {Chattopadhyay},
  \citenamefont {Singh}, \citenamefont {Manepalli}, \citenamefont {Chirila},
  \citenamefont {Yu}, \citenamefont {Walters}, \citenamefont {White},
  \citenamefont {Xiao}, \citenamefont {Tchelepi}, \citenamefont {Marcus},
  \citenamefont {Anandkumar}, \citenamefont {Hassanzadeh},\ and\ \citenamefont
  {Prabhat}}]{kashinath2021}%
  \BibitemOpen
  \bibfield  {author} {\bibinfo {author} {\bibfnamefont {K.}~\bibnamefont
  {Kashinath}}, \bibinfo {author} {\bibfnamefont {M.}~\bibnamefont {Mustafa}},
  \bibinfo {author} {\bibfnamefont {A.}~\bibnamefont {Albert}}, \bibinfo
  {author} {\bibfnamefont {J.-L.}\ \bibnamefont {Wu}}, \bibinfo {author}
  {\bibfnamefont {C.}~\bibnamefont {Jiang}}, \bibinfo {author} {\bibfnamefont
  {S.}~\bibnamefont {Esmaeilzadeh}}, \bibinfo {author} {\bibfnamefont
  {K.}~\bibnamefont {Azizzadenesheli}}, \bibinfo {author} {\bibfnamefont
  {R.}~\bibnamefont {Wang}}, \bibinfo {author} {\bibfnamefont {A.}~\bibnamefont
  {Chattopadhyay}}, \bibinfo {author} {\bibfnamefont {A.}~\bibnamefont
  {Singh}}, \bibinfo {author} {\bibfnamefont {A.}~\bibnamefont {Manepalli}},
  \bibinfo {author} {\bibfnamefont {D.}~\bibnamefont {Chirila}}, \bibinfo
  {author} {\bibfnamefont {R.}~\bibnamefont {Yu}}, \bibinfo {author}
  {\bibfnamefont {R.}~\bibnamefont {Walters}}, \bibinfo {author} {\bibfnamefont
  {B.}~\bibnamefont {White}}, \bibinfo {author} {\bibfnamefont
  {H.}~\bibnamefont {Xiao}}, \bibinfo {author} {\bibfnamefont {H.~A.}\
  \bibnamefont {Tchelepi}}, \bibinfo {author} {\bibfnamefont {P.}~\bibnamefont
  {Marcus}}, \bibinfo {author} {\bibfnamefont {A.}~\bibnamefont {Anandkumar}},
  \bibinfo {author} {\bibfnamefont {P.}~\bibnamefont {Hassanzadeh}},\ and\
  \bibinfo {author} {\bibfnamefont {n.}~\bibnamefont {Prabhat}},\ }\bibfield
  {title} {\bibinfo {title} {Physics-informed machine learning: Case studies
  for weather and climate modelling},\ }\href
  {https://doi.org/10.1098/rsta.2020.0093} {\bibfield  {journal} {\bibinfo
  {journal} {Philosophical Transactions of the Royal Society A: Mathematical,
  Physical and Engineering Sciences}\ }\textbf {\bibinfo {volume} {379}},\
  \bibinfo {pages} {20200093} (\bibinfo {year} {2021})}\BibitemShut {NoStop}%
\bibitem [{\citenamefont {Oncley}\ and\ \citenamefont
  {Horst}(2013)}]{oncley2013}%
  \BibitemOpen
  \bibfield  {author} {\bibinfo {author} {\bibfnamefont {S.}~\bibnamefont
  {Oncley}}\ and\ \bibinfo {author} {\bibfnamefont {T.}~\bibnamefont {Horst}},\
  }\href@noop {} {\emph {\bibinfo {title} {Calculation of {{Cn2}} for Visible
  Light and Sound from {{CSAT3}} Sonic Anemometer Measurements}}},\ \bibinfo
  {type} {Tech. Rep.}\ (\bibinfo {year} {2013})\BibitemShut {NoStop}%
\bibitem [{\citenamefont {Karam}\ and\ \citenamefont {Saad}(2021)}]{karam2021}%
  \BibitemOpen
  \bibfield  {author} {\bibinfo {author} {\bibfnamefont {M.}~\bibnamefont
  {Karam}}\ and\ \bibinfo {author} {\bibfnamefont {T.}~\bibnamefont {Saad}},\
  }\bibfield  {title} {\bibinfo {title} {{{BuckinghamPy}}: {{A Python}}
  software for dimensional analysis},\ }\href
  {https://doi.org/10.1016/j.softx.2021.100851} {\bibfield  {journal} {\bibinfo
   {journal} {SoftwareX}\ }\textbf {\bibinfo {volume} {16}},\ \bibinfo {pages}
  {100851} (\bibinfo {year} {2021})}\BibitemShut {NoStop}%
\bibitem [{\citenamefont {Wang}\ \emph {et~al.}(2021)\citenamefont {Wang},
  \citenamefont {Wu}, \citenamefont {Weimer},\ and\ \citenamefont
  {Zhu}}]{wang2021}%
  \BibitemOpen
  \bibfield  {author} {\bibinfo {author} {\bibfnamefont {C.}~\bibnamefont
  {Wang}}, \bibinfo {author} {\bibfnamefont {Q.}~\bibnamefont {Wu}}, \bibinfo
  {author} {\bibfnamefont {M.}~\bibnamefont {Weimer}},\ and\ \bibinfo {author}
  {\bibfnamefont {E.}~\bibnamefont {Zhu}},\ }\href@noop {} {\bibinfo {title}
  {{{FLAML}}: {{A Fast}} and {{Lightweight AutoML Library}}}} (\bibinfo {year}
  {2021}),\ \Eprint {https://arxiv.org/abs/1911.04706} {arxiv:1911.04706 [cs,
  stat]} \BibitemShut {NoStop}%
\bibitem [{\citenamefont {Vapnik}(1998)}]{vapnik1998}%
  \BibitemOpen
  \bibfield  {author} {\bibinfo {author} {\bibfnamefont {V.~N.}\ \bibnamefont
  {Vapnik}},\ }\href@noop {} {\emph {\bibinfo {title} {Statistical Learning
  Theory}}},\ Adaptive and Learning Systems for Signal Processing,
  Communications, and Control\ (\bibinfo  {publisher} {{Wiley}},\ \bibinfo
  {address} {{New York}},\ \bibinfo {year} {1998})\BibitemShut {NoStop}%
\bibitem [{\citenamefont {Molnar}(2022)}]{molnar2022}%
  \BibitemOpen
  \bibfield  {author} {\bibinfo {author} {\bibfnamefont {C.}~\bibnamefont
  {Molnar}},\ }\href@noop {} {\emph {\bibinfo {title} {Interpretable {{Machine
  Learning}}: {{A Guide}} for {{Making Black Box Models Explainable}}}}},\
  \bibinfo {edition} {second edition}\ ed.\ (\bibinfo  {publisher} {{Christoph
  Molnar}},\ \bibinfo {address} {{Munich, Germany}},\ \bibinfo {year}
  {2022})\BibitemShut {NoStop}%
\bibitem [{\citenamefont {Albertson}\ \emph {et~al.}(1995)\citenamefont
  {Albertson}, \citenamefont {Parlange}, \citenamefont {Katul}, \citenamefont
  {Chu}, \citenamefont {Stricker},\ and\ \citenamefont
  {Tyler}}]{albertson1995}%
  \BibitemOpen
  \bibfield  {author} {\bibinfo {author} {\bibfnamefont {J.~D.}\ \bibnamefont
  {Albertson}}, \bibinfo {author} {\bibfnamefont {M.~B.}\ \bibnamefont
  {Parlange}}, \bibinfo {author} {\bibfnamefont {G.~G.}\ \bibnamefont {Katul}},
  \bibinfo {author} {\bibfnamefont {C.-R.}\ \bibnamefont {Chu}}, \bibinfo
  {author} {\bibfnamefont {H.}~\bibnamefont {Stricker}},\ and\ \bibinfo
  {author} {\bibfnamefont {S.}~\bibnamefont {Tyler}},\ }\bibfield  {title}
  {\bibinfo {title} {Sensible {{Heat Flux From Arid Regions}}: {{A Simple
  Flux-Variance Method}}},\ }\href {https://doi.org/10.1029/94WR02978}
  {\bibfield  {journal} {\bibinfo  {journal} {Water Resources Research}\
  }\textbf {\bibinfo {volume} {31}},\ \bibinfo {pages} {969} (\bibinfo {year}
  {1995})}\BibitemShut {NoStop}%
\bibitem [{\citenamefont {Mahrt}(2014)}]{mahrt2014}%
  \BibitemOpen
  \bibfield  {author} {\bibinfo {author} {\bibfnamefont {L.}~\bibnamefont
  {Mahrt}},\ }\bibfield  {title} {\bibinfo {title} {Stably {{Stratified
  Atmospheric Boundary Layers}}},\ }\href
  {https://doi.org/10.1146/annurev-fluid-010313-141354} {\bibfield  {journal}
  {\bibinfo  {journal} {Annual Review of Fluid Mechanics}\ }\textbf {\bibinfo
  {volume} {46}},\ \bibinfo {pages} {23} (\bibinfo {year} {2014})}\BibitemShut
  {NoStop}%
\bibitem [{\citenamefont {Rannik}\ \emph {et~al.}(2016)\citenamefont {Rannik},
  \citenamefont {Peltola},\ and\ \citenamefont {Mammarella}}]{rannik2016}%
  \BibitemOpen
  \bibfield  {author} {\bibinfo {author} {\bibfnamefont {{\"U}.}~\bibnamefont
  {Rannik}}, \bibinfo {author} {\bibfnamefont {O.}~\bibnamefont {Peltola}},\
  and\ \bibinfo {author} {\bibfnamefont {I.}~\bibnamefont {Mammarella}},\
  }\bibfield  {title} {\bibinfo {title} {Random uncertainties of flux
  measurements by the eddy covariance technique},\ }\href
  {https://doi.org/10.5194/amt-9-5163-2016} {\bibfield  {journal} {\bibinfo
  {journal} {Atmospheric Measurement Techniques}\ }\textbf {\bibinfo {volume}
  {9}},\ \bibinfo {pages} {5163} (\bibinfo {year} {2016})}\BibitemShut
  {NoStop}%
\bibitem [{\citenamefont {Thiermann}\ and\ \citenamefont
  {Grassl}(1992)}]{thiermann1992}%
  \BibitemOpen
  \bibfield  {author} {\bibinfo {author} {\bibfnamefont {V.}~\bibnamefont
  {Thiermann}}\ and\ \bibinfo {author} {\bibfnamefont {H.}~\bibnamefont
  {Grassl}},\ }\bibfield  {title} {\bibinfo {title} {The measurement of
  turbulent surface-layer fluxes by use of bichromatic scintillation},\ }\href
  {https://doi.org/10.1007/BF00120238} {\bibfield  {journal} {\bibinfo
  {journal} {Boundary-Layer Meteorology}\ }\textbf {\bibinfo {volume} {58}},\
  \bibinfo {pages} {367} (\bibinfo {year} {1992})}\BibitemShut {NoStop}%
\bibitem [{\citenamefont {Businger}\ \emph {et~al.}(2002)\citenamefont
  {Businger}, \citenamefont {McLaren}, \citenamefont {Ogasawara}, \citenamefont
  {Simons},\ and\ \citenamefont {Wainscoat}}]{businger2002}%
  \BibitemOpen
  \bibfield  {author} {\bibinfo {author} {\bibfnamefont {S.}~\bibnamefont
  {Businger}}, \bibinfo {author} {\bibfnamefont {R.}~\bibnamefont {McLaren}},
  \bibinfo {author} {\bibfnamefont {R.}~\bibnamefont {Ogasawara}}, \bibinfo
  {author} {\bibfnamefont {D.}~\bibnamefont {Simons}},\ and\ \bibinfo {author}
  {\bibfnamefont {R.~J.}\ \bibnamefont {Wainscoat}},\ }\bibfield  {title}
  {\bibinfo {title} {Starcasting},\ }\href
  {https://doi.org/10.1175/1520-0477(2002)083{$<$}0858:S{$>$}2.3.CO;2}
  {\bibfield  {journal} {\bibinfo  {journal} {Bulletin of the American
  Meteorological Society}\ }\textbf {\bibinfo {volume} {83}},\ \bibinfo {pages}
  {858} (\bibinfo {year} {2002})}\BibitemShut {NoStop}%
\end{thebibliography}%

\end{document}